\documentclass[prd,nofootinbib,showpacs,twocolumn]{revtex4-1}

\usepackage{amstext,amsmath,amssymb,amsfonts,bbm}
\usepackage[latin1]{inputenc}
\usepackage{graphicx}
\usepackage{hyperref}
\usepackage{amsthm}
\usepackage{tocvsec2}
\usepackage{enumerate}
\usepackage{subfigure}
\usepackage{color}

\usepackage{hyperref}

\def\beq{\begin{equation}}
\def\be{\begin{equation}}
\def\ee{\end{equation}}
\def\bes{\begin{eqnarray}}
\def\ees{\end{eqnarray}}

\def\f{\frac}

\theoremstyle{definition}
\theoremstyle{definition}
\theoremstyle{definition}
\theoremstyle{definition}
\theoremstyle{definition}
\theoremstyle{definition}

\begin{document}
\maxtocdepth{subsection}
%%%%%%%%%%%%%%%%%%%%%%%%%%%%%%%%%%%%%%%%%%%%%%%%%%%

\title{\large \bf Unruh effect without trans-horizon entanglement}
\author{{Carlo Rovelli}}\email{smerlak@cpt.univ-mrs.fr}
\author{{Matteo Smerlak}}\email{rovelli@cpt.univ-mrs.fr}
\affiliation{Centre de Physique Th\'eorique, Campus de Luminy, Case 907, 13288 Marseille Cedex 09, France}

\date{\small\today}

%%%%%%%%%%%%%%%%%%%%%%%%%%%%%%%%%%%%%
\begin{abstract}\noindent
We estimate the transition rates of a uniformly accelerated Unruh-DeWitt detector coupled to a quantum field with reflecting conditions on a boundary plane (a ``mirror").  We find that these are essentially indistinguishable from the usual Unruh rates, viz. that the Unruh effect persists in the presence of the mirror.  This shows that the Unruh effect (thermality of detector rates) is not merely a consequence of the entanglement between left and right Rindler quanta in the Minkowski vacuum. Since in this setup the state of the field in the Rindler wedge is pure, we argue furthermore that the relevant entropy in the Unruh effect cannot be the von Neumann entanglement entropy. We suggest, in alternative, that it is the Shannon entropy associated with Heisenberg uncertainty.

\end{abstract} 
%%%%%%%%%%%%%%%%%%%%%%%%%%%%%%%%%%%%%%

%MSC numbers: 81T45 (principal), 57M20, 81T25, 83C45 (secondary)
%\keywords{}
%\tableofcontents

%%%%%%%%%%%%%%%%%%%%%%%%%%%%%%%%%%%%%%
\maketitle

\section{Introduction}

An accelerated particle detector clicks even in the vacuum. This is not surprising \emph{per se}: the detector receives energy from whichever device is accelerating it, and there is no reason why this energy should not be exchanged with the field. What is surprising, however, is the \emph{thermal} character of these transitions in the case of uniform acceleration, discovered by Unruh \cite{Unruh1976}: thermal states are states of maximal entropy---whence the entropy of ``acceleration radiation"?

An oft-heard explanation \cite{Wald1994} of this puzzle traces this remarkable effect back to the (earlier) Fulling-Davies \emph{thermalization theorem} \cite{Fulling1973,Davies1975}: the Minkowski vacuum state of a free quantum field consists of entangled pairs of left and right Rindler quanta,\footnote{A ``left Rindler mode" is a solution of the field equation that vanishes on the right Rindler wedge \cite{Wald1994}.} with Boltzmann-like coefficients. The argument goes as follows. Due to its acceleration, a detector in the (say) right Rindler wedge is causally disconnected from the left modes of the field, which vanish everywhere on its trajectory. Hence it does not interact with the full \emph{pure} vacuum state of the field, but only with the \emph{mixed} state obtained by tracing out the left degrees of freedom screened by the Rindler horizon. This mixed state is thermal by the thermalization theorem. For instance, Jacobson writes \cite{Jacobson}
\begin{quote}
``The essence of the Unruh effect is the fact that the density matrix describing the Minkowski vacuum, traced over the states in the region $z < 0$, is precisely a Gibbs state for the boost Hamiltonian at a ``temperature" $T=1/2\pi$."
\end{quote}
In this light, the entropy of the Unruh radiation appears to be related to the von Neumann entropy of the improper mixture of right Rindler quanta \cite{Israel1976,Bombelli1986}.\footnote{See \cite{Bekenstein1994} for more references on this idea of \emph{entanglement entropy}.}

If by ``Unruh effect" one means the thermal character of the vacuum \emph{field} fluctuations observable within a Rindler wedge, this is clearly correct.  But if we restrict the attention just to the detector's transition rates, and by ``Unruh effect" one means---as in Unruh's original work and as we do here---the thermal character of the \emph{detector}'s transition rates, then, we argue here, the story is subtler and there is more to learn.  

A difficulty with the entanglement interpretation of the Unruh effect (in the sense specified) has been pointed out repeatedly, e.g. in \cite{Sriramkumar1996,Schlicht2004,Akhmedov2008}: it violates causality. The Rindler horizon of an accelerated observer depends on its entire worldline, with proper time ranging from minus to plus infinity. But a physical effect cannot depend on the future history of the observer. This motivated Schlicht to study the Unruh effect in causal terms \cite{Schlicht2004}; he concluded 
\begin{quote}
``A detector which is asymptotically at rest for $t\rightarrow\pm\infty$, is moving for an arbitrarily long (but finite) time with almost uniform acceleration and perceives an (almost) thermal radiation-spectrum. Because he returns from asymptotic rest to asymptotic rest, there is no acceleration horizon in this case, and no hidden degree of freedom."
\end{quote}

Here we present a far stronger argument against the reduction of the Unruh effect to a mere consequence of the thermalization theorem. We show that an accelerated detector continues to measure the Unruh temperature also in the right Minkowski \emph{half-space} $\{ z>0\}$, with reflecting boundary conditions at $z=0$.  In this setup there are no left modes in the first place and the detector interacts with all the degrees of freedom of a pure state of the quantum field. If the Unruh effect were just a consequence of the statistical nature of the state obtained by tracing over the left modes, one would expect that the detector transition rates would fail to be thermal in this setup.  On the contrary, we show that the detector records the same Unruh thermal radiation as without mirror. 

This result shows that the relation between acceleration and temperature discovered by Unruh is more general than that determined by the entanglement entropy due to the screening of a causal horizon: it holds also for pure states of the field.  In the last section, we speculate on the possible nature of the entropy relevant in the general case.

To avoid any misunderstanding, let us stress again that by ``Unruh effect'' we mean the thermality of detector transition rates for uniform acceleration in the vacuum.  Stronger statements are sometimes implied in the literature (the thermalization theorem itself, Planckian transition rates, non-trivial Bogoliubov coefficients, etc.); this ambiguity has been a source of confusion in the past.\footnote{Recall the debates on the statistics inversion in odd dimensions noted by Takagi \cite{Takagi1986} and on the triviality of Bogoliubov coefficients for circular motion \cite{Letaw1980,Denardo1978}.} Here we choose to stick to Unruh's original operational definition of the effect. 

The transition rates of a particle detector in the presence of a plane boundary have been discussed by several authors \cite{Pringle1989,Ohnishi1992,Langlois2006}. To our knowledge, however, none of them has drawn what we consider the salient conclusion of this calculation: it shows that the Unruh effect is \emph{not} just about entanglement with hidden degrees of freedom.

In this paper, the signature of spacetime is $(-+++)$, and we use geometrized units where $c=\hbar=1$. Our main reference on the Unruh effect is Takagi's thorough review \cite{Takagi1986}. 

\section{The causal Unruh effect}

To study the response of a particle detector in the vacuum in a causal manner, we need the concept of \emph{instantaneous transition rate} \cite{Svaiter1992,Schlicht2004}. Let $x(\tau)$ be a timelike worldline parametrized by its proper time $\tau$, describing the motion of a pointlike monopole (Unruh-DeWitt) detector in the quantum vacuum. Assume the detector is turned on at time $\tau_0$. The number of transitions per unit time between states with energy difference $\omega$ at time $\tau$ is, at first order in perturbation theory, proportional to
\be\label{defrate}
R_{\tau,\tau_0}(\omega)=2\ \textrm{Re}\int_0^{\tau-\tau_0}ds\ e^{-i\omega s}w(\tau,\tau-s),
\ee
where $w(\tau,\tau-s)=W\big(x(\tau),x(\tau-s)\big)$ is the (positive frequency) Wightman $2$-point function\footnote{The $2$-point function is really a bi-distribution with extended singularities. We refer to the papers of Louko and his students, e.g.\,\cite{Langlois2006} and \cite{Louko2008}, for up-to-date discussions of regularization issues.} of the field evaluated along the detector's worldline. When $\omega>0$, the transition corresponds to the \emph{absorption} of a quantum $\omega$ by the detector; when $\omega<0$, it corresponds to spontaneous \emph{emission} of the same quantum.

Consider now a detector which, from the moment $\tau_0$ at which it is turned on, undergoes a uniform acceleration with magnitude $a$. The causal version of the Unruh effect states that, when $\tau-\tau_0\gg a^{-1}$, the rates \eqref{defrate} reach a limit $R_a(\omega)$ which is \emph{thermal} at the inverse temperature $\beta_a=2\pi/a$ \cite{Svaiter1992}. This means that the \emph{detailed balance condition} holds:
\be\label{detailedbalance}
\f{R_a(\omega)}{R_a(-\omega)}=e^{-\beta_a\omega}.
\ee
Note that \emph{no assumption regarding the detector's trajectory before $\tau_0$ or later than $\tau$ is made}. In particular, the Rindler horizon plays no direct r\^ole.\footnote{Note also that the condition \eqref{detailedbalance} does \emph{not} imply that $R_a(\omega)$ assume the Planckian form they would have if detector were inertial and immersed in a thermal bath with temperature $\beta_a$. Indeed, for a massive Klein-Gordon field, it does not: in the large mass limit $m\rightarrow\infty$, the Unruh rates read instead \cite{Takagi1986}
\be\label{massiveunruhrates}
R_a(\omega)\sim \f{a}{8\pi}e^{-\beta_am/\pi}e^{-\beta_a\omega/2}.
\ee}

This said, it is true, of course, that during the period from $\tau_0$ to $\tau$ when it is switched on, the detector only has a partial access to the field degrees of freedom. This is easy to see by going to the instantaneous rest frame of the detector at the switching time $\tau_0$. There, the worldline for $\tau\geq\tau_0$ has coordinates
\be\left\{ \begin{array}{ll}\label{hyperbolic}
t(\tau)&=a^{-1}\sinh\big(a(\tau-\tau_0)\big)\\z(\tau)&=a^{-1}\cosh\big(a(\tau-\tau_0)\big).
\end{array} \right.
\ee
The hypersurface $C_0=\{t=0\}$ is a Cauchy surface for the field.  But it is apparent from \eqref{hyperbolic} that at any time $\tau>\tau_0$, the past light-cone of $x(\tau)$ only contains a subset of $C_0^>=C_0\cap\{z> 0\}$, the part of $C_0$ on the right of $z=0$. That is, as long as it is accelerated, the detector is connected to only \emph{half} of the Cauchy data on $C_0$. In this sense the acceleration ``screens" some of the field's degrees of freedom---and motivates the partial trace over $\{z<0\}$ mentioned by Jacobson.

\section{Unruh effect with a mirror}

Let us now change the setup by placing a \emph{mirror} (Dirichlet boundary conditions\footnote{It is immediate to adapt the following discussion to Neumann boundary conditions.}) at $z=0$, so that the field vanishes for $z\leq 0$. In this case, $C_0$ is not a Cauchy surface any longer; but its right part $C_0^>$ is.  And \emph{all the points of $C_0^>$ are causally connected with the accelerated trajectory \eqref{hyperbolic}}. Does the Unruh effect  disappear?\footnote{Before preforming this calculation we were not sure what to expect and we asked many of the best experts in the field: most of them answered, incorrectly, that they expected no thermal detector's response ``because there can be no thermal detector's response in a pure state".}

By the method of images the Wightman function $W^>$ in the presence of the mirror is related to the usual one $W$, namely without the mirror, by
\be\label{mirrorwightman}
W^>(x,y)=W(x,y)-W(x,Ry),
\ee
where $Ry$ is the mirror image of $y$ with respect to the $z=0$ plane. Now, by Poincar\'e invariance, we know that $W$ only depends on the invariant interval $(x-y)^2$
\be
W(x,y)=w\big((x-y)^2\big).
\ee 
Locality implies furthermore that $W(x,y)\rightarrow0$ when $x$ and $y$ are spacelike separated and $(x-y)^2\rightarrow\infty$. This is the case of the points $x(\tau)$ and $Rx(\tau-s)$ along the accelerated trajectory \eqref{hyperbolic}: their squared proper distance $\lambda_\tau(s)=\big(x(\tau)-Rx(\tau-s)\big)^2$ is given by
\be\label{distance}
\lambda_\tau(s)=2a^{-2}\big(1+\cosh a(2\tau-s)\big),
\ee
where we set $\tau_0=0$ for notational simplicity. On the interval $0\leq s\leq\tau$, we have
\be\label{distancebound}
\lambda_\tau(s)\geq a^{-2}e^{a\tau}, 
\ee 
and thus the boundary term $W\big(x(\tau),Rx(\tau-s)\big)$ quickly vanishes as $\tau\gg a^{-1}$.

More specifically, if we assume that $w$ has a power-law decay at large spacelike separations, viz.
\be
\vert w(\lambda)\vert\leq \lambda^{-k},
\ee
for some $k>0$, then we have 
\be\label{expbound}
 \big | W\big(x(\tau),Rx(\tau-s)\big) \big |\leq a^{2k}e^{-ka\tau}.
\ee
Thus, when evaluated on the accelerated worldline, \emph{the Wightman function \eqref{mirrorwightman} is essentially indiscernible from its value in empty space}, viz. without the mirror.\footnote{Note that, if the decay of $w$ is exponential instead of power-law (e.g. for a massive field), this difference is even more negligible: it is exponentially-exponentially vanishing.} See Fig. 1.

%\begin{widetext}

\begin{figure}[t]
\centering
\subfigure[\ $W\big(x,x(\tau_0)\big)$]{
\includegraphics[scale=.47]{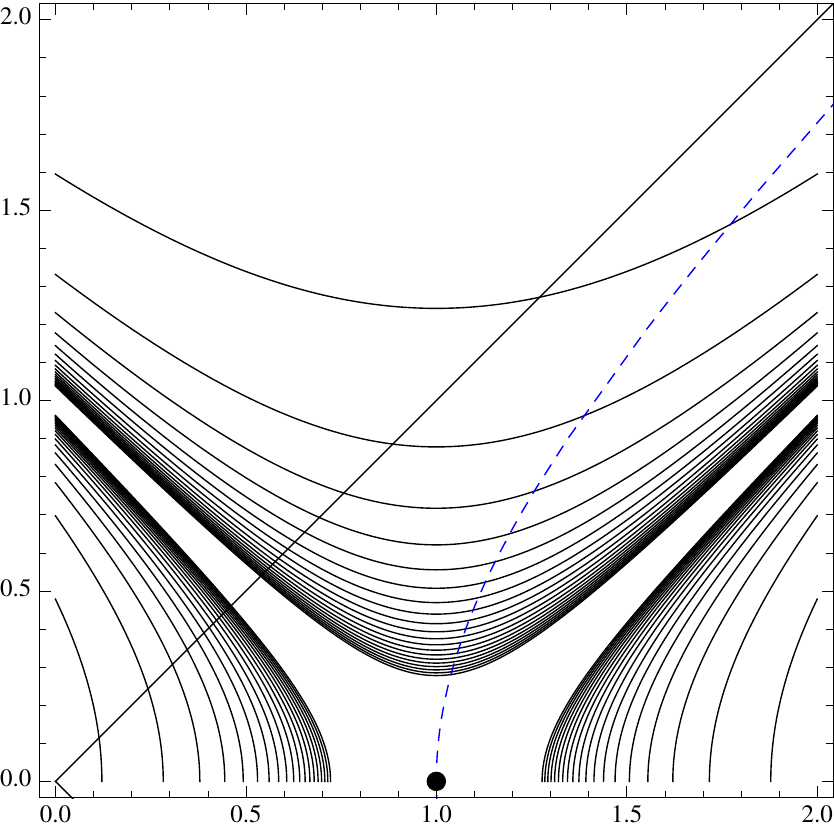}
}
\subfigure[\ $W^>\big(x,x(\tau_0)\big)$]{
\includegraphics[scale=.47]{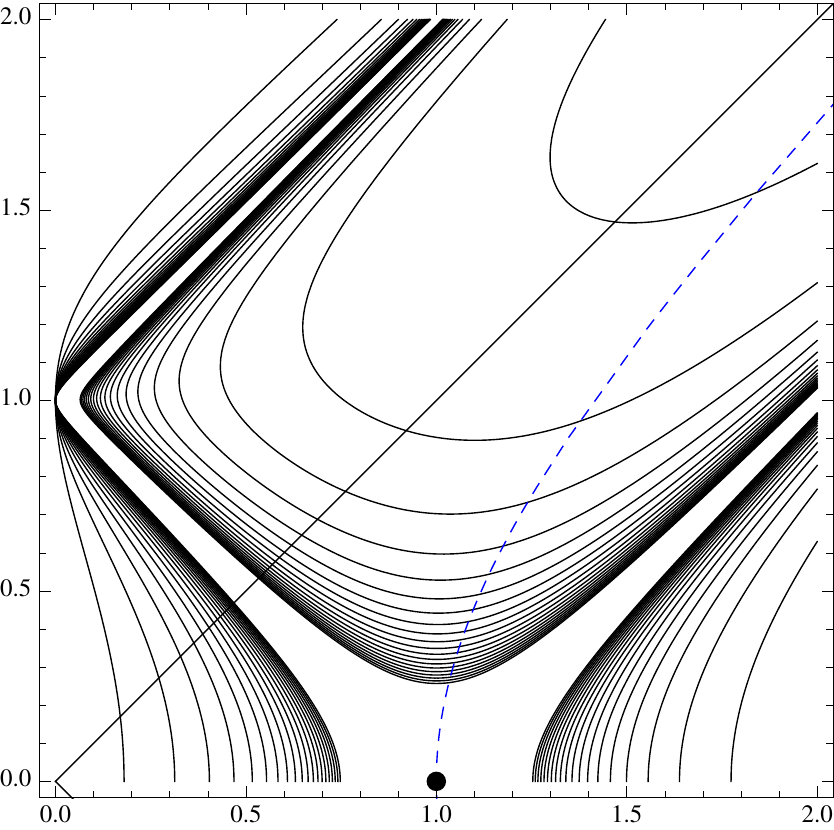}
}

\caption{Wightman functions in the $(z,t)$ plane, for a massless scalar field in empty space (a) and with a mirror at $z=0$ (b). The dashed hyperbolae represent uniformly accelerated trajectories $x(\tau)$ with $a=1$, the dots indicate the switching time $\tau_0$, and the solid line $\{z=t\}$ is the boundary of the causal past of $x(\tau)$ as $\tau\rightarrow\infty$.}
\end{figure}
%%\end{widetext}

From \eqref{defrate} and \eqref{mirrorwightman}, the detector transition rates in the presence of the mirror read
\be
R^>_{\tau}(\omega)=R_\tau(\omega)+B_\tau(\omega)
\ee
where $R_\tau(\omega)$ is the empty space term (with the Unruh limit $R_a(\omega)$ when $a\tau\rightarrow\infty$), and
\be\label{deferrorterm}
B_\tau(\omega)=-2\ \textrm{Re}\int_0^{\tau} ds\ e^{-i\omega s}W\big(x(\tau),Rx(\tau-s)\big)
\ee
is the boundary term. Using \eqref{expbound}, it is immediate to see that, as $a\tau\rightarrow\infty$,
\be
B_\tau(\omega)=\mathcal{O}\big(\tau e^{-ka\tau}\big).
\ee
Thus, the thermal limit
\be
R_\tau^>(\omega)\underset{a\tau\rightarrow\infty}{\longrightarrow} R_a(\omega)
\ee
is reached also with the mirror. 
\emph{The absence of hidden field degrees of freedom does not spoil the Unruh effect.}

\section{Choice of the mirror's frame}

An ingredient we used to arrive at this conclusion is the fact that, when the detector is switched on (at time $\tau_0$), the mirror is \emph{at rest} relative to the latter. What difference would it make if the mirror were moving towards the detector at this instant,  $(dz/dt)(\tau_0)<0$? The answer is that the transition rates $R^>_\tau(\omega)$ would start converging to their Unruh limit $R_{a}(\omega)$ only after a transient regime, during which the mirror recorrelates the field with itself.

%\emph{not} match the Unruh ones $R_a(\omega)$. 

Indeed, denote $\tau_1$ the time when the detector most approaches the mirror,  $(dz/dt)(\tau_1)=0$. Then the function $\lambda_\tau(s)$ measuring the proper distance between $x(\tau)$ and the image $Rx(\tau-s)$ reads
\be
\lambda_\tau(s)=2a^{-2}\Big(1+\cosh a\big(2(\tau-\tau_1)-s\big)\Big)
\ee
instead of \eqref{distance}. If $\tau_{1}\leq\tau\leq2\tau_1$, $\lambda_\tau(s)$ has a minimum at $s^*=2(\tau-\tau_1)$, given by $\lambda_\tau(s^*)=4a^{-2}$. This minimum violates the exponential bound \eqref{distancebound}. Hence, during this regime, the corrective term $B_\tau(\omega)$ does not decrease. Physically, this is because the mirror correlates the field \emph{after} the u-turn of the detector with its value \emph{before} the u-turn. The difference between $W\big(x(\tau),x(\tau_0)\big)$ and $W^>\big(x(\tau),x(\tau_0)\big)$ for $\tau_{1} \leq\tau\leq2\tau_1$ is clearly visible on Fig. 2.

%The importance of the choice of the mirror's rest frame was not acknowledged in earlier works on the Unruh effect in the presence of boundaries \cite{PhysRevD.39.2178,TakagiOhnishi,Langlois:2005nf}. 

\begin{figure}
\centering
\subfigure[\ $W\big(x,x(\tau_0)\big)$]{
\includegraphics[scale=.47]{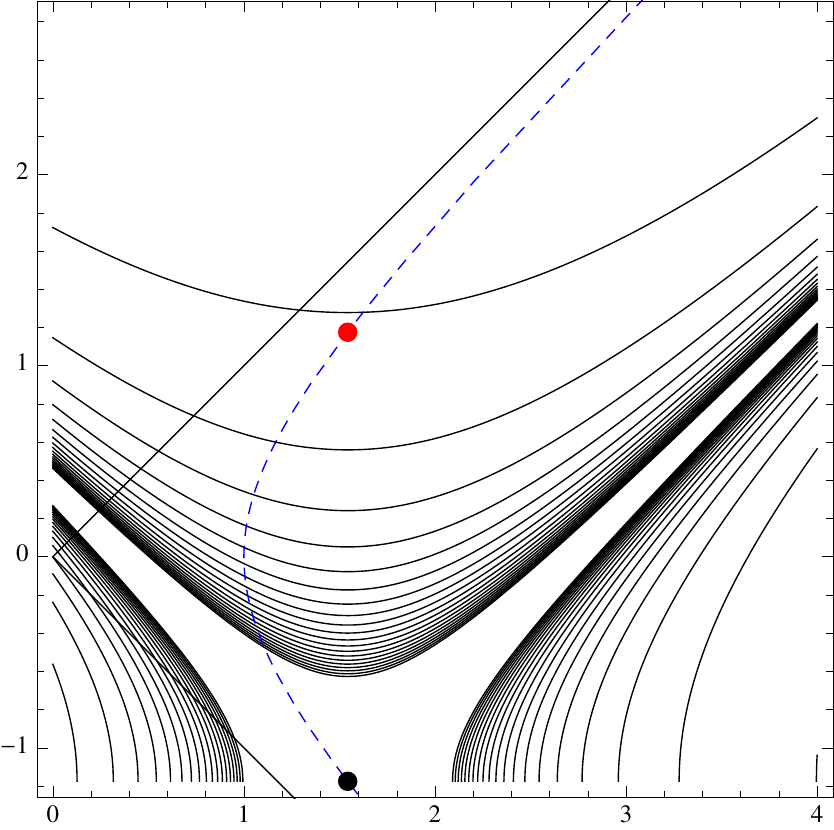}
}
\subfigure[\ $W^>\big(x,x(\tau_0)\big)$]{
\includegraphics[scale=.47]{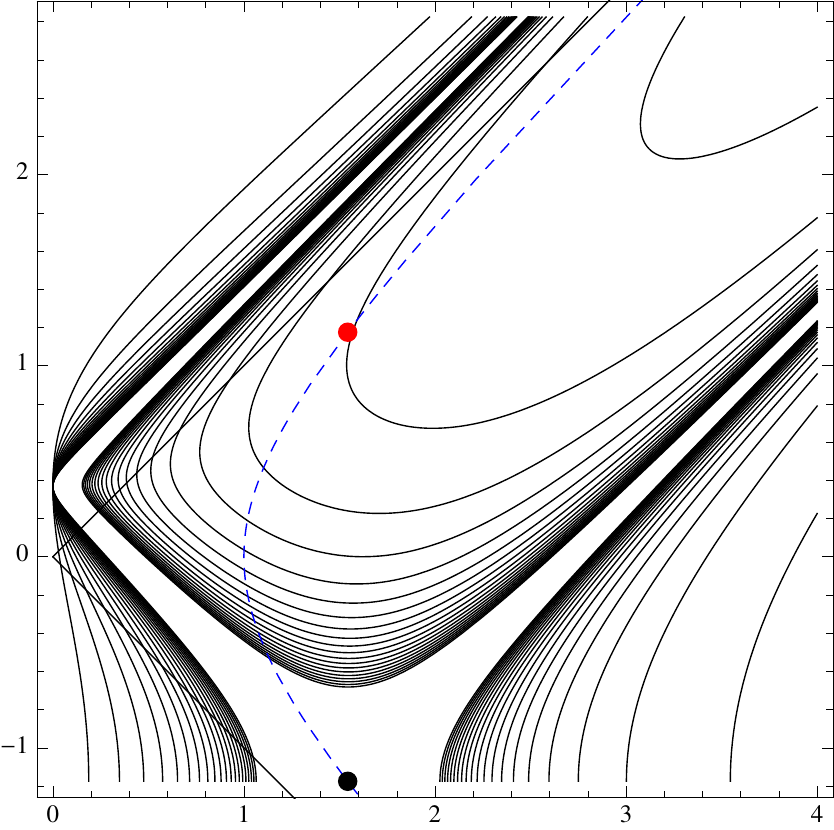}
}

\caption{Wightman functions in the $(z,t)$ plane, for a massless scalar field in empty space (a) and with a mirror at $z=0$ which is now \emph{moving} towards the detector when $\tau=\tau_0$ (b). The black dots (bottom) indicate the switching time $\tau_0$, and the red dots (up) the end of the transient regime $2\tau_1$.}
\end{figure}
%%\end{widetext}

\section{Discussion and outlook}

The fact that the Unruh effect persists in the presence of a mirror on the edge of the detector's horizon does not mean that the state of the quantum field in the right region $\{z>0\}$ is the same with and without the mirror. It is not.  In the presence of the mirror, the state is pure; in the absence of the mirror, it is mixed.  Observables different from detector rates can distinguish between the two cases.  What our result shows is, instead, that the thermal character of the transition rates of a uniformly accelerated detector, pointed out by Unruh, cannot be reduced to the effect of tracing out modes behind the horizon; it is present also with a pure state.

What makes this result puzzling---and therefore interesting---is that it blatantly contradicts a basic tenet of statistical physics: the connection between thermality and ignorance of the microstate. In normal systems, indeed, temperature arises as a consequence of microscopic fluctuations, which prevent us from knowing the energy of the system. For this reason, we are led to describe it by means of a probability distribution over its spectrum of states, namely the one which is least biaised towards a specific state (given the macroscopic constraints) \cite{Jaynes1957}. The extent to which this description is uncertain is measured by the corresponding Gibbs (classical) or von Neumann (quantum) entropy. As the temperature is lowered down to zero, and the system is driven to its fundamental state, this uncertainty decreases and the entropy goes to zero. 

In the standard setup of the Unruh effect, that is without the mirror, the Fulling-Davies thermalization theorem appears to confirm this scenario: it relates the thermal response of the accelerated detector to the ignorance of the state of the field in the left wedge due to its entanglement accross the edge of the wedge. The entropy associated to the Unruh effect can thus be tentatively identified as the von Neumann entanglement entropy. As recalled in the introduction, this insight is often considered a key ingredient to understand the nature of the elusive Bekenstein-Hawking black hole entropy. 
 
Here, however, we are forced into a radically different interpretation of the Unruh effect. Since the state of the field in the presence of the mirror is pure, indeed, we cannot account for the thermality of the detector rates by invoking the ignorance of the microstate of the system. If its energy is uncertain to the accelerated detector, then, it must be \emph{for another reason}, and the entropy measuring this uncertainty must be different from the von Neumann entropy. What is this reason, and what is the corresponding entropy? Besides clarifying the nature of the Unruh effect, we believe that understanding this point will shed new light on the problem of black hole entropy.

Without entering a full discussion, we present here a simple speculative answer.  Any state  
$|\psi\rangle$ of a quantum system defines a probability distribution
\be
           p_{\psi}(a)=|c(a)|^2=|\langle a|\psi\rangle|^2
\ee
on the spectrum of any observable $A$ of the system.  Here $|a\rangle$ is the eigenstate of $A$ with eigenvalue $a$,   
\be
|\psi\rangle=\sum_a c(a)|a\rangle
\ee
and we assume for simplicity a discrete and non degenerate spectrum.  This probability distribution has a Shannon entropy 
\be
           S_\psi(A)= - \sum_a p_\psi(a) \ln p_\psi(a)
\ee
which measures the intrinsic uncertainty in the outcome of the measurement of $A$ in the state $|\psi\rangle$  due to Heisenberg uncertainty relations \cite{Wehner2010}.  

Imagine we double the degrees of freedom of the system. Then we can associate to $|\psi\rangle$ the state 
\be
           |\Psi\rangle=\sum_a c(a)|a\rangle \otimes |a\rangle'
\ee
where $|a\rangle'$ are the states of the second copy of the system. If we trace over the degrees of freedom of this second copy of the system we obtain the mixed state 
\be
           \rho =\sum_a p(a)|a\rangle\langle a|
\ee
which has a von Newmann entropy equal to the Shannon entropy of $p_\psi$ and which is such that the probability distribution of the outcomes of $A$ measurements is the same as for $\psi$ (this is not true for other observables, obviously).

Consider this in the case of an accelerated thermometer.  An (ideal) inertial thermometer is essentially a device that couples to the energy of the field.  In a thermal situation, its response depends on the probability distribution of energy eigenstates.  The inertial vacuum state  $|0\rangle$  is an eigenstate of the energy $H$ with vanishing eigenvalue and therefore an inertial thermometer measures a vanishing temperature.  However, an accelerated detector does not couple to the energy operator $H$ (which generates translations in Minkowski time), but rather to the restriction $K^>$ of the boost generator $K$ to the right Rindler modes\footnote{Notice that $K$ does not mix left and right modes, hence $K^>$ is well defined.}, which generates translations in proper time along its accelerated trajectory.  The two operators $H$ and $K^>$ do not commute, and the vacuum is an eigenstate of $K$ but not of $K^>$.  Let $|k\rangle$ be the eigenstates of $K^>$ with eigenvalue $k$ and let 
\be
           p_0(k)=|\langle 0|k\rangle|^2
\ee
be the probability that the accelerated detector measures the $k$ eigenvalue in the inertial vacuum (again, assume discrete spectrum and non degeneracy for simplicity).    Then the measurement outcomes are characterized by a Shannon entropy 
\be 
           S_{0}(K^>)= - \sum_k p_0(k) \ln p_0(k).
\ee 
Such Shannon entropy is present, and in fact is the same, with or without a mirror. 

In the case with the mirror, it measures the uncertainty due to the quantum fluctuations associated to measuring an observable ($K^>$) in a state ($|0\rangle$) which is not its eigenstate. In the case without the mirror, this happens to correspond also to the entanglement entropy precisely as the example above: the eigenstates of $K^>$ are correlated with modes beyond the horizon, so that the entanglement entropy in $\rho$ is equal the Shannon entropy of the pure state. The main point is that the probability distribution $p_0(k)$, which is the quantity of interest from the detector point of view, is independent on whether or not there are modes behind the horizon to be traced over. What is relevant is only the non-commutativity between $H$ and $K^>$, which determines the probability distribution $p_0(k)$. 

Hence, we suggest that the entropy responsible for the relation between acceleration and temperature is not the von Neumann entropy of the statistical state obtained by tracing beyond horizon modes, but rather the Shannon entropy deriving from the Heisenberg uncertainty associated to the observable measured by the accelerated detector. The latter is a measurement of both quantum and thermal uncertainty, with no \emph{a priori} distinction between the two. This is concomitant with Smolin's suggestion that quantum and thermal fluctuations could be indistinguishable in a gravitational context where a unique algebra of observables at fixed time is in general non defined \cite{Smolin1986}.  

%The precise relation between this idea and the presence of horizons, which appears to have a strict relation with the thermal character of vacuum fluctuations, will be studied elsewhere. 

%The reason why the ``hidden modes" perspective on the Unruh effect has remained tantalizing is its remarkable unifying power: it relates the thermal features of quantum radiation in Rindler spacetime to that of a black hole or the deSitter universe, and indeed to the one observed any spacetime with a bifurcate Killing horizon\footnote{Note that in the deSitter case, no acceleration is involved.}; moreover, it complies with Hawking's original intuition that black hole radiation is related to the capture of particles by the horizon, and seems concomitant with the r\^ole played by the Bekenstein-Hawking entropy in the generalized second law.
\medskip

\paragraph*{Acknowledgements.}

C.R. thanks Stefan Hollands for a useful conversation on the Unruh effect.

%%%%%%%%%%%%%%%%%%%%%%%%%%%%%%%%%%%%%%%%%%%%%%%%%%%
%%%%%%%%%%%%%%%%%%%%%%%%%%%%%%%%%%%%%%%%%%%%%%%%%%%
\bibliographystyle{apsrev4-1}

%\bibliography{library}

%merlin.mbs apsrev4-1.bst 2010-07-25 4.21a (PWD, AO, DPC) hacked
%Control: key (0)
%Control: author (72) initials jnrlst
%Control: editor formatted (1) identically to author
%Control: production of article title (-1) disabled
%Control: page (0) single
%Control: year (1) truncated
%Control: production of eprint (0) enabled
%

%%%%%%%%%%%%%%%%%%%%%%%%%%%%%%
%%%%%%%%%%%%%%%%%%%%%%%%%%%%%%

\end{document}